\documentclass[10pt,prl,twocolumn,showpacs]{revtex4}
\pdfoutput=1
\usepackage{graphicx,amsmath,amssymb}
\usepackage{hyperref}
\begin{document}

\def\nn{\nonumber}
\def\kc#1{\left(#1\right)}
\def\kd#1{\left[#1\right]}
\def\ke#1{\left\{#1\right\}}
\renewcommand{\Re}{\mathop{\mathrm{Re}}}
\renewcommand{\Im}{\mathop{\mathrm{Im}}}
\renewcommand{\b}[1]{\mathbf{#1}}
\renewcommand{\c}[1]{\mathcal{#1}}
\renewcommand{\u}{\uparrow}
\renewcommand{\d}{\downarrow}
\newcommand{\bsigma}{\boldsymbol{\sigma}}
\newcommand{\blambda}{\boldsymbol{\lambda}}
\newcommand{\Tr}{\mathop{\mathrm{Tr}}}
\newcommand{\sgn}{\mathop{\mathrm{sgn}}}
\newcommand{\sech}{\mathop{\mathrm{sech}}}
\newcommand{\diag}{\mathop{\mathrm{diag}}}
\newcommand{\Pf}{\mathop{\mathrm{Pf}}}
\newcommand{\half}{{\textstyle\frac{1}{2}}}
\newcommand{\sh}{{\textstyle{\frac{1}{2}}}}
\newcommand{\ish}{{\textstyle{\frac{i}{2}}}}
\newcommand{\thf}{{\textstyle{\frac{3}{2}}}}
\newcommand{\SUN}{SU(\mathcal{N})}
\newcommand{\N}{\mathcal{N}}
\newcommand{\be}{\begin{equation}}
\newcommand{\ee}{\end{equation}}
\newcommand{\tr}{\mathop{\mathrm{tr}}}
\newcommand{\sign}{\mathop{\mathrm{sign}}}

\title{A Holographic Fractional Topological Insulator}

\author{Carlos Hoyos}
\author{Kristan Jensen}
\author{Andreas Karch}

\affiliation{
Department of Physics, University of Washington, Seattle, WA
98195-1560, USA}

\date\today

\begin{abstract}
We give a holographic realization of the recently proposed low energy effective action describing a fractional topological insulator. In particular we verify that the surface of this hypothetical material supports a fractional quantum Hall current corresponding to half that of a Laughlin state.
\end{abstract}


\maketitle

\emph{Introduction.}---The concept of a fractional topological insulator (FTI) was recently introduced in~\cite{Maciejko:2010tx}. Time reversal ($T$) invariant topological insulators (TI) have been a very active field of recent research ~\cite{Qi2010,Moore2009,Hasan2010}. The properties of a TI
can be described at different levels of microscopic detail. For non-interacting band insulators in $3+1$ dimensions, topological band theory can identify a $\mathbb{Z}_2$ topological invariant~\cite{Kane2005,Fu2007,Moore2007}. The relevant low-energy dynamics of a generic band insulator can be modeled by a single massive Dirac fermion. In a $T$-invariant theory the mass is real and its sign becomes the $\mathbb{Z}_2$ invariant, this is the language we find most useful in this work. Integrating out the massive fermion yields a topological field theory (TFT)~\cite{Qi2008}. The TFT has a Lagrangian proportional to $\theta\vec{E}\cdot\vec{B}$, where $\theta$, which naively vanishes by $T$-invariance, may take the values $0$ or $\pi$ after accounting for the proper quantization of magnetic flux. The value of $\theta$ is given by the phase of the mass of the fermion per the Adler-Bell-Jackiw~(ABJ) anomaly~\cite{ABJ1969}.
Physically, this leads to a single massless Dirac cone on the surface between a TI and an ordinary insulator (or vacuum). Thus, in an electric field the surface supports a Hall current with a conductivity corresponding to half that of an integer Hall state at filling fraction $1$.

As with the quantum Hall effect, one expects the picture to get modified in the presence of interactions. The proposal of~\cite{Maciejko:2010tx} describes a potential low energy theory describing a FTI, that is a state supporting on its surface a fractional quantum Hall current with effective filling fraction $1/2m$.
All one needs to assume is that the electron fractionalizes into $m$ partons of charge $1/m$ (in units of the electron charge $e$). A statistical ``color'' gauge field is also added so that physical states have an electric charge given by an integer multiple of $e$.
From this construction the fractional quantum Hall current follows immediately via the ABJ anomaly. A question that was not completely resolved in~\cite{Maciejko:2010tx} is whether the color gauge field has to be in a confined or a deconfined phase. A confined phase would have the advantage of being completely gapped. As emphasized in~\cite{Swingle:2010rf} such a phase is potentially problematic. The authors of~\cite{Swingle:2010rf} proved a theorem that states that $\theta$ cannot be fractional in a completely gapped theory unless the ground state on $T^3$ is degenerate, which differs from the proposal of~\cite{Maciejko:2010tx} for a confined theory. The basic problem can be seen both at the level of a theory with gauge fields and partons as well as for the TFT obtained after integrating out the fermions. In the TFT, flux quantization allows fractional theta as long as one accounts for both magnetic and color magnetic fluxes. However if the color gauge fields confine, we expect their magnetic fluxes to be screened and not to take on quantized values. So the Dirac quantization argument only holds in the deconfined phase. Including the partons, the ABJ anomaly of the axial symmetry can be used to derive a fractional $\theta$. However, the axial symmetry is typically broken dynamically in the confined phase. So one should take the color gauge field to be deconfined.

While the system with deconfined color gauge fields is gapless it still describes an insulator. The only charged fields, the partons, are gapped. The gauge fields are better thought of as phonons; they contribute to thermodynamics and mediate interactions between the gapped partons. However they do not directly contribute to electric transport. Now we need a color theory in a deconfined phase to study. Non-Abelian gauge theories in $3+1$ dimensions typically confine with a few charged matter fields, but can give deconfined phases with enough additional matter.
As long as the extra matter fields are electrically neutral, we still describe a gapped spectrum of charge carriers interacting via a gapless phonon bath. Since these theories are often strongly coupled it is hard to establish the phase realized by the gauge theory. In this letter we explicitly demonstrate that this model of a FTI
can be realized via holography~\cite{Maldacena:1997re,Gubser:1998bc,Witten:1998qj}. We take our phonon bath to be ${\cal N}=4$ super Yang-Mills theory (SYM) with a large number $m$ of colors at strong `t Hooft coupling. In this limit the theory has a dual description in terms of type IIB supergravity on $AdS_5 \times S^5$. We add electrically charged partons via D7 probe branes~\cite{Karch:2002sh}. We show that this system realizes a Hall current on an interface with a Hall conductivity corresponding to the filling fraction $1/2m$, just as predicted in~\cite{Maciejko:2010tx}.

\emph{Axial anomaly.}---In order to calculate the Hall conductivity in our system, we review the basic anomaly argument of~\cite{Maciejko:2010tx}.
Take the simplest microscopic model for a TI: a complex Dirac fermion $\psi$, its charge conjugate $\bar{\psi}=\psi^{\dagger}\gamma^0$, and a real action
\be
\label{freeaction}
{\cal L} = \bar{\psi} (i \partial_{\mu} \gamma^{\mu} - M) \psi.
\ee
The theory is $T$-symmetric only if the mass $M$ is real. For $M=0$, axial rotations are also a symmetry
\be
\psi \rightarrow e^{-i \phi \gamma_5/2} \psi .
\ee
In the massive theory they shift the phase of the mass
\be
 M \rightarrow e^{-i \phi} M,
\ee
so one can always use the explicitly broken axial symmetry to rotate $M$ to be real and positive.

In an interacting quantum theory this axial rotation is often anomalous. For example, consider a Dirac fermion transforming in some representation $R$ of a gauge group $G$. Define an index $C$ of the representation via
\be
\text{Tr}_R T^a T^b = C \delta^{ab}.
\ee
For the (fundamental + anti-fundamental) representation of $G=SU(m)$ this gives $C=1$, whereas for a charge $q$ (in units of $e$) representation of $G=U(1)$ we get $C=q^2$.  Now while the classical gauge theory is invariant under axial rotations, the quantum theory is not. The path integral picks up an extra phase from the Jacobian which can written as a shift in the action by
\be
\Delta {\cal L} = C e^2\frac{\phi }{32 \pi^2} \epsilon_{\mu \nu \rho \sigma} F^{\mu \nu} F^{\rho \sigma},
\ee
which effectively shifts $\theta$ as
\be
\theta \rightarrow \theta - C \phi.
\ee
This is the famous ABJ anomaly. After integrating out a heavy fermion of mass $M$ and phase $\phi$, the theory remembers $\phi$ as a change in the effective $\theta$ angle.

For applications to electro-magnetism we choose to start with a $T$-invariant $U(1)$ gauge theory that has no $\theta$-angle in vacuum. All contributions to $\theta$ come from integrating out heavy fermions. The effective $\theta$ after integrating a heavy fermion of charge $e$ is either $0$ when the mass is positive or $\pi$ when
the mass is negative. A 4d theory with a mass that passes through zero corresponds to an interface between topologically trivial and non-trivial insulators. $M(x)$ having a root guarantees the existence of a massless mode localized on the interface.

To get a FTI, we consider a theory of $m$ partons of charge $1/m$ in units of $e$. 
The index $C$ is now
\be C = m \cdot (1/m)^2 = 1/m. \ee
If the partons have a negative mass, we generate an effective $\theta = \pi/m$, leading to a fractional quantum Hall conductance corresponding to the filling fraction
\be
\nu = \frac{1}{2m}
\ee
on an interface between positive and negative mass regions.

This calculation holds for the theory of a single ${\cal N}=2$ supersymmetric hypermultiplet interacting with an $\N=4$ SYM phonon bath. Simply vary the hypermultiplet mass from real and positive to real and negative. The interface carries a Hall current corresponding to $\nu=1/2m$ if we assign the partons charge $1/m$ as above. We find it more convenient, for the purpose of counting powers of $m$ in the large $m$ limit that underlies our holographic calculation, to assign charge $1$ to the parton, giving the electron a total charge $m$. The anomaly argument then predicts an effective
filling fraction
\be
\nu = \frac{m \cdot 1^2}{2} =\frac{m}{2}.
\ee
Note that this still corresponds to the same fractional quantum Hall state and is merely a matter of convention.

\begin{figure*}[htc]
\begin{center}
\includegraphics[width=16cm]{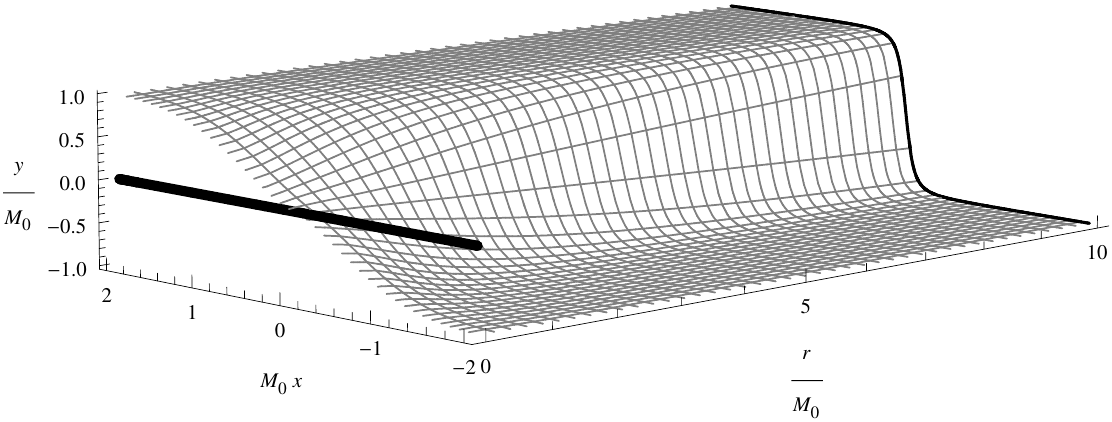}
\end{center}
\caption{\label{thePlot}We plot a numerically generated brane embedding dual to an interface between a trivial insulator and a FTI. The mass profile in the field theory is roughly a step function at $x=0$ with a height $M_0$.  The D7 brane intersects the D3 branes (indicated by the thick black line at $\rho=y=0$) at the interface there.
}
\end{figure*}

\emph{Holographic Calculation.}---Holographically, $m$ partons in the fundamental representation of $SU(m)$ can be added to the ${\cal N}=4$ SYM phonon bath dual to supergravity on $AdS_5 \times S^5$ by embedding a D7 brane~\cite{Karch:2002sh} wrapping an $S^3$ inside the $S^5$ and extending in all $5$ directions of $AdS_5$.
In addition to minimizing its worldvolume, the brane has a Wess-Zumino (WZ) term in its action. This couples the $U(1)$ gauge field on the brane to the form fields present in the background, in particular to the $m$ units of D3-brane flux that support the background geometry. These terms will be crucial in what follows.
Writing the background metric as
\be
g = (r^2+Y^2)g_{3,1} +\frac{dr^2+r^2d\Omega_3^2+dY^2+Y^2d\phi^2}{r^2+Y^2}.
\ee
The brane embedding is given by functions $Y(r),\phi(r)$. For a mass $M$ hypermultiplet, $Y=|M|$, $\phi=\text{arg }M$ is
an exact solution. For this solution the 3-sphere shrinks to zero size at $r=0$ and so the brane terminates there, above the bottom of AdS. To realize a domain wall we want an embedding where $M$ is a real function of one of the spatial coordinates, say $x$, with a root. For every function $M(x)$ there is a unique brane embedding $Y(r,x)$.

Being due to an anomaly, the Hall current entirely comes from the WZ term in the probe brane action. Using the conventions of \cite{Davis:2008nv} we have
\begin{align}
\nonumber
S_{WZ}&= - \frac{1}{2} (2 \pi \alpha')^2 T_7 \int A \wedge F \wedge G_5, \\
\nonumber
\int_{S^5} G_5 &= 2 \kappa_{10}^2 T_3 m, \quad \quad
T_3 T_7 = \frac{1}{g^2 (2 \pi)^4 (2 \pi \alpha')^6} ,\\
\label{conventions}
2 \kappa_{10}^2 &= g^2 (2\pi)^3 (2 \pi \alpha')^4
\end{align}
The D7 brane worldvolume can be parametrized with the coordinates $t,y,z$ along the interface, the angles on the wrapped 3-sphere, $\phi$, and a new coordinate $\theta$ defined by $\theta=\arctan\,Y/r$. These last five coordinates describe the 5-sphere by
\begin{equation}
d\Omega_5^2=d\theta^2+\cos^2\theta d\Omega_3^2+\sin^2\theta d\phi^2,
\end{equation}
so that the WZ term then simply becomes
\be
S_{WZ} = - \frac{T_7 (2 \pi \alpha')^2}{2}  \int (A \wedge F)_{tyz} \, \left(\int_{S^5} G_5\right) \frac{\Delta \phi}{2\pi}
\ee
where the $A\wedge F$ term only includes components along the $t,y,$ and $z$ directions and where in the last factor we used that the fraction of the full $S^5$ wrapped by the D7 is $\Delta \phi/2\pi$. We define $\Delta\phi$ to be the range of $\phi$ realized over the embedding. For a $T$-invariant FTI we have $\Delta \phi = \pi$, but the expression is valid for any $\Delta \phi$. Using Eq.~(\ref{conventions}) we see that there is a net induced Chern-Simons (CS) term
\be
S_{WZ} = -\frac{m}{4 \pi} \frac{\Delta \phi}{2 \pi} \int (A\wedge F)_{tyz}.
\ee
For a $T$-invariant embedding where $\phi$ changes discontinuously at some $x_0$, the CS term is localized at $x_0$. As in~\cite{Davis:2008nv}, it corresponds to a half-integer CS term of level
\be
k= m/2
\ee
inducing a Hall current with $\nu=m/2$ as predicted from the anomaly argument. In fact, close to the zero crossing we can think of our D7 as intersecting the D3 orthogonally, as already pointed out in~\cite{Ryu:2010hc}. In this case our brane embedding actually becomes identical to the one in~\cite{Davis:2008nv}. While the latter is unstable and requires a UV completion to make sense of the condensed phase, in our case the single fundamental Dirac cone on the interface is protected by the topology of the 4d bulk FTI.

\emph{Example.}---We showed that, independent of the details of the embedding, the Hall current takes on the correct quantized value just determined by the change in phase of the mass. It would be desirable to at least construct one such embedding. Unlike most brane embeddings considered in the literature, one complication here is that we need to solve a partial differential equation in two variables, $x$ and $r$. In particular, we consider embeddings that interpolate between the masses $-M_0$ and $M_0$. We also account for the phase $\phi$ by letting $Y$ be negative for $\phi=\pi$ and positive for $\phi=0$. Thus for $x \rightarrow \pm \infty$, $Y$ approaches $\pm M_0$. Moreover the curve $Y(\infty,x)$ maps to $M(x)$ in the field theory and can be chosen freely. The Lagrangian for the brane is
\be
{\cal L} = r^3 \sqrt{1 + (\partial_{r} Y)^2 + (r^2+Y^2)^{-2} (\partial_x Y)^2}.
\ee
Notably, this Lagrangian exhibits a scaling symmetry under which $Y,r\rightarrow \xi Y,\xi r$ and $x\rightarrow x/\xi$.

We solve the corresponding equation of motion with two different methods. First, we construct an analytic solution for a particular $M(x)$ in a series expansion with the interface at $x=0$. We also choose to study antisymmetric solutions about the interface. Since at large enough $r$ the linearized equation for $Y$ is accurate, our solutions will be indexed with a single parameter $M_0$ that scales as $M_0\rightarrow \xi M_0$. The scaling symmetry then suggests an ansatz $Y(r,x)=M_0 f_0(xr)$. Indeed an $f_0$ exists so that this ansatz solves the equation of motion for $Y$ to $O(M_0^3)$, with a full solution of the form
\be
\label{seriesSol}
Y(x,r)=M_0\sum_{n=0}^{\infty}(M_0 x)^{2n}f_n(xr).
\ee
The first term in the series is
\be
\label{seriesLead}
Y(x,r)=\frac{M_0xr}{\sqrt{1+(xr)^2}}+O(M_0^3x^2).
\ee
This solution is accurate for $M_0x\ll 1$ and $r\gg M_0$; away from these limits the higher order terms compare to the leading one. Near the defect, the mass profile is then a step function, $M(x)= \text{sign}(x)M_0+O(M_0^3x^2)$. This embedding also encodes the value of the chiral condensate $c=\langle\bar{\psi}\psi\rangle+\ldots$ in the field theory; terms not explicitly displayed contain the superpartners of $\psi$. From dimensional analysis alone, the condensate (which has to vanish as $M_0$ vanishes) can be written as
\be
c(x)=\text{sign}(x)\frac{M_0}{x^2}\sum_{n=0}^{\infty}c_n(M_0x)^{2n}.
\ee
Indeed, the condensate can be measured from the $r^{-2}$ coefficient of the embedding at large $r$, giving $c(x)=\text{sign}(x)M_0/4x^2+O(M_0^3x^2)$.

Not only does the series solution Eq.~(\ref{seriesSol}) fail at small $r$, but none of its terms satisfy the correct small$-r$ boundary condition. This is that the brane ends smoothly at $r=0$, which implies that $\partial_r Y(0,x)=0$. For $Y(0,x)$ nonzero, this implies that at small $r$ the embedding is $Y(x,r)\sim Y_0(x)+Y_2(x)r^2+\ldots$. When $Y_0(x)$ has a root, as our solutions do, there is a non-analyticity in the solution at the root and $r=0$. In lieu of this difficulty, we also obtain an embedding numerically.

\begin{figure}[t]
\begin{center}
\includegraphics[scale=0.75]{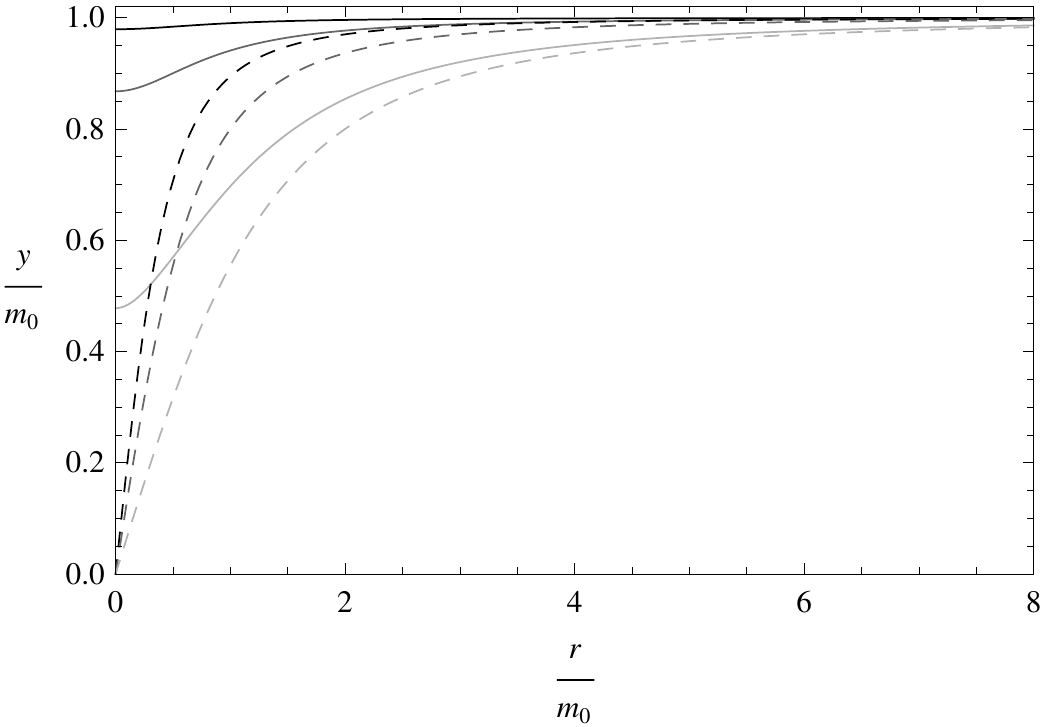}
\end{center}
\caption{
\label{comparePlot}We compare the numerical embedding in Fig.~\ref{thePlot} (solid) with the series solution Eq.~\ref{seriesLead} (dashed) at three values of $x$. The dark lines indicate $x=2/M_0$, the gray lines $x=4/3M_0$, and the light lines $x=2/3M_0$. The solutions agree down to $r\sim 5 M_0$.
}
\end{figure}

We do so by employing a heat method. The minimal area action ensures that if we take the ``time'' derivative of a field configuration $Y$ to be proportional to the variation of the action, it will quickly settle to a correct solution. We choose to find a numerical completion to our series solution, and so we impose the boundary condition that it matches the leading term of the series Eq.~(\ref{seriesSol}) at a large $r_c\gg M_0$. The profile here is almost a step function. We also impose that the embedding is constant at large $x_c\gg 1/M_0$ as well as the smoothness condition at $r=0$. The resulting solution will be dual to the theory with a mass $M(x)$ that is close to a step function.

We generated a solution with the parameters $r_c=20M_0$ and $x_c=4/M_0$. We plot a portion of it in Fig.~\ref{thePlot}. At large $r$ the profile approximates a step function and at small $r$ the embedding has a single root at the interface, asymptoting to the constant embedding far away from the interface over a distance of roughly $1/M_0$. Also, we compare this embedding with the series solution Eq.~(\ref{seriesLead}) at several values of $x$ in Fig.~(\ref{comparePlot}).

This work was supported in part by DOE grant DE-FG02-96ER40956.

\bibliography{holofti3d}

\end{document}